\documentclass[a4paper,fleqn]{cas-sc}
\usepackage[numbers]{natbib}
\usepackage{booktabs}
\usepackage{adjustbox}
\usepackage{longtable}
\usepackage{caption}
\usepackage{mathtools}
\usepackage{amssymb}
\usepackage{amsmath}
\usepackage{etoolbox}

\newlength{\myalignwidth}

\def\tsc#1{\csdef{#1}{\textsc{\lowercase{#1}}\xspace}}
\tsc{WGM}
\tsc{QE}
\tsc{EP}
\tsc{PMS}
\tsc{BEC}
\tsc{DE}
\begin{document}
\let\WriteBookmarks\relax
\def\floatpagepagefraction{1}
\def\textpagefraction{.001}
\let\printorcid\relax

\shorttitle{}

\shortauthors{Chaoqian Wang {\it et~al.}}

\title [mode = title]{When greediness and self-confidence meet in a social dilemma}                      

\author[1]{Chaoqian Wang}
\ead{CqWang814921147@outlook.com}
\credit{Conceptualization; Methodology; Writing}

\author[2]{Wenqiang Zhu}
\credit{Methodology; Validation}

\author[3]{Attila Szolnoki}
\cormark[1]
\cortext[cor1]{Corresponding author}
\ead{szolnoki.attila@ek-cer.hu}
\credit{Conceptualization; Validation; Writing}

% Address/affiliation
\address[1]{Department of Computational and Data Sciences, George Mason University, Fairfax, VA 22030, USA}
\address[2]{Institute of Artificial Intelligence, Beihang University, Beijing 100191, China}
\address[3]{Institute of Technical Physics and Materials Science, Centre for Energy Research, P.O. Box 49, H-1525 Budapest, Hungary}

% Here goes the abstract
\begin{abstract}
A greedy personality is usually accompanied by arrogance and confidence. This work investigates the cooperation success condition in the context of biased payoff allocation and self-confidence. The first component allows the organizer in a spatial public goods game to receive a different proportion of goods than other participants. The second aspect influences the micro-level dynamics of strategy updates, wherein players can maintain their strategy with a certain weight. Analytical results are obtained on square lattices under the weak selection limit. If the organizer attempts to monopolize the public goods, cooperation becomes more attainable. If the confidence increases, cooperation is inhibited. Consequently, these elements have conflicting effects on cooperation, and their simultaneous presence can result in a heterogeneous change of the critical synergy factor. Our theoretical findings underscore the subtle implications of a mutual trait that may manifest as greediness or self-confidence under different circumstances, which are validated through Monte Carlo simulations.
\end{abstract}

% Use if graphical abstract is present
% \begin{graphicalabstract}
% \includegraphics{figs/grabs.pdf}
% \end{graphicalabstract}

% Research highlights
\begin{highlights}
\item Examining biased allocation and self-confidence in spatial public goods game
\item Calculating cooperation success conditions in weak selection limit
\item Conflicting effects yield a non-monotonic critical synergy factor
\item Analytical results validated via Monte Carlo simulations
\end{highlights}


\maketitle

\section{Introduction}\label{secintro}
The dynamism of various facets of reciprocity---be they direct, indirect, or network reciprocity---have been unequivocally demonstrated to wield significant influence over system behaviors, particularly when there is a need to sustain costly cooperation among self-interested, or more crudely put, selfish agents~\cite{nowak_s06}. These mechanisms, chiefly concerned with pairwise interactions among players, have been observed to incorporate higher-order interactions~\cite{perc_jrsi13,sigmund_10}. The public goods game (PGG) is an illustrative example of such complex interactions, involving simultaneous decision-making processes through multi-body or group interactions~\cite{wang_jw_pla22,xiao_sl_epjb22,wang2022reversed,hua_sj_csf3}. Players may opt to contribute or abstain from contributing to a common pool, reaping the benefits of the overall contributions regardless of their individual decisions. In a spatial population, where players engage in limited yet enduring interactions with others, reciprocity manifests on an additional level~\cite{szolnoki_pre09c,yu_fy_csf22,wang2021public,wang2022between,wang2023public,wang_cq_c23,xie_k_csf23,ding_r_csf23}. Here, the intricate web of relations among agents means a player is not limited to a single game, but finds themselves immersed in several others. A pragmatic approach for a player would be to partake in the group where they serve as the central agent, encircled by proximate neighbors. Concurrently, said player also engages in games instigated by their neighbors. Consequently, a player positioned on a node with a $k$ degree finds themselves partaking in $G=k+1$ PGGs. This setup could potentially underpin a reciprocal mutual aid system which promotes a degree of cooperation.

Assuming the most rudimentary scenario where players consistently maintain their strategies across all the games they participate in and disregard strategy diversity~\cite{zhang_cy_epl10}, there still exists considerable flexibility in the implementation of a realistic model. To elaborate, groups do not necessarily correspond to a player, who may be more incentivized to invest effort in a venture they have personally initiated. Such dedication could be recognized and appreciated by the others. This could be simply expressed by allocating enhanced contributions in a biased manner. Specifically, a $0\leq w_L \leq 1$ fraction of the total income is allotted to the central player while the remaining $1-w_L$ is distributed among the participating neighbors. The $w_L=1/G$ scenario represents the traditional PGG model, where the income is equally distributed among all participants. The $w_L=0$ limit corresponds to the situation where the central player allocates all income to the neighbors. While this may initially seem irrational, there have been empirical studies indicating the existence of similar practices in certain tribes where partners generally offer a larger share to an associate in an ultimatum game, signaling their honest intentions~\cite{henrich_aer01}. The other extreme case, $w_L=1$, denotes that the central player retains all the benefits. Interestingly, even this seemingly greedy scenario can reflect a cooperative intent and represent a form of mutual aid~\cite{nowak_sa95,allen2013spatial,su2018understanding}. One can contemplate a barn constructed by an entire Amish community, yet later solely utilized by a single farmer. This study aims to explore the potential ramifications when players exhibit a specific $w_L$ value.

The unequal distribution of collective benefits has previously been the subject of extensive investigation~\cite{zhang_hf_pa12,cong_r_epl16,szolnoki_amc20,wang_q_amc18,su2018understanding,bin_l_amc23}. For instance, how income is allocated remains a central issue in the ultimatum game~\cite{guth_jebo82,sigmund_sa02,szolnoki_prl12,wang_xf_srep14,chen_w_epl15,szolnoki_epl12}. For the current study, however, the diverse allocation within a group comprising several participants is of greater relevance. In certain scenarios, the individual portion accrued by a participant can be strongly contingent on their investment capability~\cite{fan_rg_pa17}. Additionally, the heterogeneous interaction topology is a critical aspect where income allocation is proportional to an agent's weight (degree) in the graph~\cite{peng_d_epjb10}. In more sophisticated model configurations, players possess an extra skill and keep track of their previous round earnings~\cite{meloni_rsos17}. Yet, our current model is straightforward, emphasizing the fundamental element of biased allocation. For example, it can be applied to regular graphs where players have equal-sized neighborhoods, thus participating in an equal number of joint groups. Moreover, we presuppose homogeneous players who behave similarly and apply a pre-established allocation policy in each case. This characteristic could prove to be crucial, as it has been widely observed that a heterogeneous population, wherein players are unequal, could serve as a mechanism that encourages cooperation~\cite{perc_pre08,santos_n08}.

Players may differ in their views about their groups, and their approach to strategies can also be distinct. For example, they may show reluctance to alter their existing strategies, a phenomenon explained from various perspectives. This could be a result of a specific cost related to change~\cite{szabo_prl02}, or it could be interpreted as a form of self-confidence~\cite{li_k_srep16,szolnoki_pre18,wang2023evolution}. This strategy change inertia or updating passivity has been identified as a separate mechanism that significantly influences the evolutionary process~\cite{szolnoki_pre09,liu_rr_epl10,zhang_yl_pre11,wang2023inertia,wang2023conflict}. To quantitatively track this effect, we introduce a $0\leq w_R \leq 1$ weight parameter, which determines the likelihood of retaining the original strategy during the elementary dynamical process. At $w_R=0$, this effect is completely absent, and we revert to the traditional death--birth rule~\cite{ohtsuki_jtb06}. In the opposite extreme, when $w_R=1$, there is no proper evaluation because all agents adamantly stick to their original strategy, despite the theoretical cooperation success condition equating to the birth-death rule as $w_R\to 1$~\cite{wang2023inertia}. In between these extremes, at $w_R=1/G$ where $G$ denotes the group size, the strategy of the central player and the strategies of the neighbors carry equal weight and we revert to the imitation rule~\cite{wang2023inertia,nowak_n04b}.

This work simultaneously considers the aforementioned effects within the framework of PGG, with players situated on a square lattice. It is important to note that the biased allocation, which can also be interpreted as autocratic behavior, and the indifference towards alternative players representing diverse strategies, may stem from a shared trait. If an individual exhibits higher levels of autocracy and retains more public goods when they organize a group, it may also display traits of arrogance, meaning they have a high self-regard and are not prone to learning from others' strategies. Therefore, the weight factors representing these traits can be similar in size. Moreover, all the mentioned details of the proposed model are strategy-neutral, making it unclear whether they support cooperation or not. Specifically, we assume the analytically feasible weak selection limit, where payoff values merely slightly alter the reproductive fitness of competing strategies.

Our main goal is to determine the critical synergy factor for the success of cooperation based on the control parameters and to uncover the consequences of their simultaneous presence. In the next section, we will define our model, and our primary findings will be presented in Section~\ref{sec_theo}. Monte Carlo simulations were also conducted to validate and confirm our theoretical results. The comparisons will be presented in Section~\ref{sec_nume}. Our primary conclusions are summarized in Section~\ref{sec_conclu}, where potential implications will also be discussed.

\section{Model}\label{sec_model}
In the study of spatial population dynamics, the model utilizes an $L\times L$ square lattice with periodic boundary conditions. Hence, the total population $N=L^2$. Each individual, referred to as an agent, inhabits a vertex on the lattice and forms a group of $G=k+1$ members, comprising of itself and $k$ of its neighbors. Consequently, each agent partakes in $1+k$ groups, either organized by itself or by its neighbors. The group formed by agent $i$ is represented by $\Omega_i$. Consequently, the collection of agent $i$'s neighbors can be expressed as $\Omega_i\setminus \{i\}$. The common choice of group size is $G=5$ ($k=4$, von~Neumann neighborhood) or $G=9$ ($k=8$, Moore neighborhood).

During each elementary Monte Carlo step, a random agent $i$ is selected to update its strategy $s_i$ based on the payoff acquired from participating in the public goods games. Specifically, agent $i$ organizes a public goods game within its group $\Omega_i$. Each participant $j\in\Omega_i$ contributes a cost $c>0$ to the group if cooperating ($s_j=1$) or contributes nothing if defecting ($s_j=0$). The combined investments of all participants $\sum_{j\in\Omega_i}s_j c$ is amplified by a synergy factor $r>1$ to generate the public goods, which are then distributed among group members.

Distinct from the conventional public goods game where the goods are evenly distributed, this study extends this notion by allowing the potential for uneven distribution between the organizer and other players. Specifically, the organizer is allotted a portion $w_L$ ($0\leq w_L\leq 1$), while the remaining players are evenly allocated the remaining proportion $1-w_L$; that is, each of the other players receives $(1-w_L)/k$. Hence, as the organizer, agent $i$ receives a payoff of $w_L r\sum_{j\in\Omega_i}s_j c-s_i c$ from group $\Omega_i$. Correspondingly, agent $i$ also participates in groups organized by its neighbors $g\in\Omega_i\setminus\{i\}$, receiving a payoff in those groups as a standard player. The payoff of agent $i$ is the average over the $k+1$ groups, calculated by:
\begin{equation}\label{eq_payoff}
    \pi_i=\frac{1}{k+1}
    \left\{
    \left(w_L r\sum_{j\in\Omega_i}s_j c-s_i c\right)
    +
    \sum_{g\in\Omega_i\setminus\{i\}}\left(\frac{1-w_L}{k}r\sum_{j\in\Omega_g}s_j c-s_i c\right)
    \right\}.
\end{equation}

As underscored, Eq.~(\ref{eq_payoff}) broadens the traditional public goods game by incorporating the self-allocation parameter $w_L$. At $w_L=0$, all public goods are allocated to the other players, while at $w_L=1$, all public goods are allocated to the organizer. At $w_L=1/G$, the public goods are distributed equally, reducing Eq.~(\ref{eq_payoff}) to the traditional public goods game scenario.

In alignment with previous studies~\cite{wang2023inertia,mcavoy2020social}, the payoff $\pi_i$ is transformed to fitness $F_i=\exp{(\delta \pi_i)}$, where $\delta\to 0^+$ is a weak selection strength limit. Therefore, a strategy with a higher fitness has a marginal advantage to reproduce more frequently. To calculate the strategy updating probability, we also compute the payoff of agent $i$'s neighbors and convert them to fitness in a similar manner. Consequently, the strategy of agent $i$ is replaced by the strategy of an agent $j\in\Omega_i$ with probability $W(s_i\gets s_j)$, which is defined by the generalized death--birth rule~\cite{wang2023evolution},
\begin{equation}\label{eq_DB}
    W(s_i\gets s_j)
    =
    \begin{cases} 
    \displaystyle{\frac{(1-w_R)/k\cdot F_j}{w_RF_i+(1-w_R)/k\cdot \sum_{\ell\in \Omega_i\setminus\{i\}}F_{\ell}}},  & \mbox{if $j\neq i$,} \\
    \displaystyle{\frac{w_R F_j}{w_RF_i+(1-w_R)/k\cdot \sum_{\ell\in \Omega_i\setminus\{i\}}F_{\ell}}}, & \mbox{if $j=i$.}
    \end{cases}
\end{equation}

In Eq.~(\ref{eq_DB}), $\sum_{j\in\Omega_i}W(s_i\gets s_j)=1$ is normalized. Eq.~(\ref{eq_DB}) extends the traditional death--birth rule~\cite{ohtsuki_jtb06} by introducing a self-learning weight $w_R$, following a similar logic to self-allocation. The agent $i$ learns the strategy of agent $j$ proportional to the fitness in the group $\Omega_i$, taking self-learning into consideration. The case of $j=i$ implies that agent $i$ does not learn the strategy from others. At $w_R=0$, Eq.~(\ref{eq_DB}) reduces to the traditional death--birth rule, where the fitness of agent $i$ is disregarded. At $w_R=1/G$, Eq.~(\ref{eq_DB}) simplifies to the imitation rule, where the fitness of agent $i$ is compared equally with all neighbors. An elementary Monte Carlo step concludes once the randomly selected agent $i$ in the system updates its strategy. A full Monte Carlo step encompasses $N$ elementary steps, ensuring that the strategy of each agent is updated on average once.

\begin{figure}
	\centering
		\includegraphics[width=.7\textwidth]{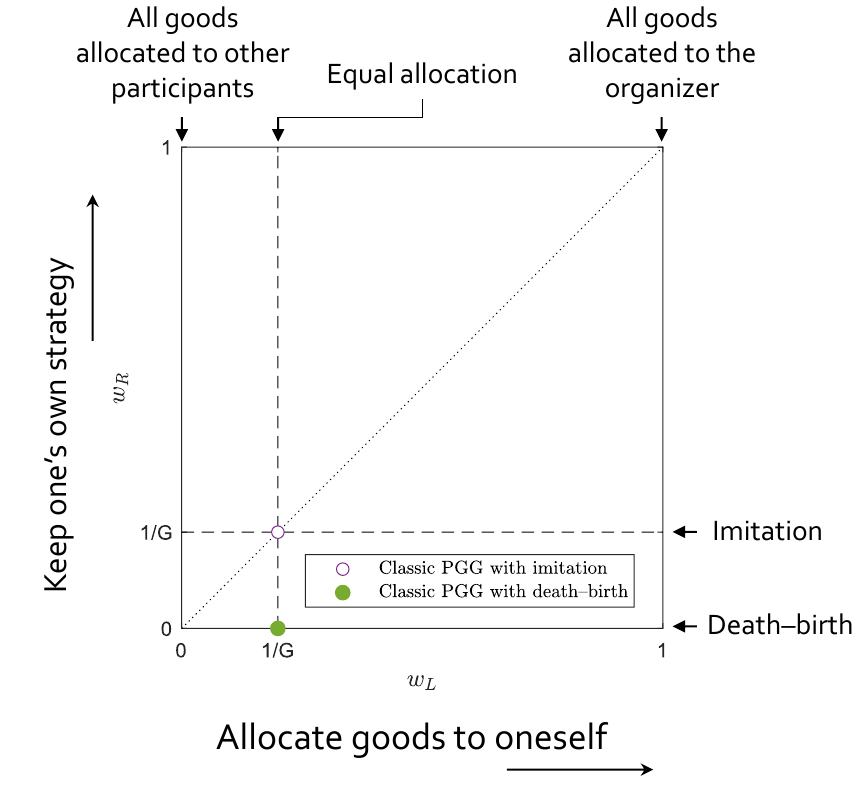}
	\caption{Comprehensive parameter plane of the extended model, where weight factors $w_L$ and $w_R$ determine the degree of biased allocation and the reluctance to change strategy, respectively. Critical values of $w_L$ and $w_R$, including $0$, $1/G$, and $1$, are highlighted on the axis. As indicated by the legend, open and closed circles represent positions of the traditional public goods game using imitation and death--birth microscopic strategy updating dynamics, respectively. The dotted diagonal line represents the trajectory where both weight factors are varied simultaneously.} 
	\label{fig_Diagram}
\end{figure}

Our model's key parameters are the weight factors, $w_L$ and $w_R$, which dictate the bias in allocation and the rate of self-learning, respectively. In Fig.~\ref{fig_Diagram}, we unveil the comprehensive parameter plane, highlighting the important weight values. These values have particular implications. When $w_L=1$, the total earnings from the communal pool are allocated solely to the focal player. Conversely, when $w_L=0$, every participant benefits from the pool while the focal player gains nothing. The midway scenario of $w_L=1/G$ recaptures the traditional public goods game (PGG) where all group members equally share the proceeds from the common pool. Shifting our attention to the other weight factor, $w_R=0$ signifies the classic death--birth dynamics, where the new strategy of the focal player is exclusively drawn from the strategies of the neighbors. When $w_R=1/G$, all strategies present in the group are potential candidates in equal measure, which aligns with the well-established imitation rule. Finally, in the limit where $w_R \to 1$, players tenaciously cling to their current strategies, thereby causing the evolution to stagnate. On the parameter plane, we also demarcate with a dotted line the trajectory where both weight factors are simultaneously altered. This trajectory represents the typical system behavior when both the effects of biased allocation and self-confidence are operative in the extended model with equal weights.

In the ensuing section, we explore and analyze how the critical synergy factor for cooperation success evolves in the presence of these skewed allocations and self-confidence biases.

\section{Theoretical analysis}\label{sec_theo}
We assume that the evolutionary process begins from a state with the presence of $N_C$ cooperative players. In essence, the initial proportion of cooperation is $N_C/N$. When the selection strength, denoted as $\delta$, equals zero, the system defaults to the dynamics of the voter model~\cite{clifford1973model}. In this state, cooperation will ultimately dominate the entire population with a probability of $\rho_C=N_C/N$~\cite{cox1983occupation,cox1986diffusive}. Consequently, under a minimal selection strength of $\delta\to 0^+$, if $\rho_C>N_C/N$, selection leans towards cooperation, which implies that evolution promotes the success of cooperative behavior. Here, $\rho_C$ can be gauged by the average final proportion of cooperation obtained from independent runs.

Our objective in Section~\ref{sec_condi} is to pinpoint the condition that enables the success of cooperation, while Section~\ref{sec_conflict} focuses on exploring the inherent features of this condition.

\subsection{The condition for cooperation success}\label{sec_condi}
To discern the requisite condition for cooperation success, we utilize the identity-by-descent (IBD) method~\cite{su2018understanding,allen2014games}. Initially, we introduce $n$-step random walks. Fundamentally, this refers to moving to a random neighbor during each $1$-step random walk. The quantity after completing $n$-step walks is represented as $x^{(n)}$, where $x$ could be $\pi$, $F$, and $s$. The $x^{(n)}$ quantity is indistinguishable among various agents since the square lattice is a vertex-transitive graph, where an agent cannot identify its location by examining the network structure.

Based on the random walks' definition, we can rewrite the payoff calculation in Eq.~(\ref{eq_payoff}) to obtain an agent's expected payoff from $n$ steps away, as described in Eq.~(\ref{eq_piwalk}),
\begin{align}
	\pi^{(n)}&=\frac{1}{k+1}\left\{\left(w_L r(k s^{(n+1)}+s^{(n)})c-s^{(n)}c\right)+k\left(\frac{1-w_L}{k} r(k s^{(n+2)}+s^{(n+1)})c-s^{(n)}c\right)\right\}\nonumber
	\\
	&=\left(\frac{w_L}{k+1}r-1\right)s^{(n)}c+\frac{1+(k-1)w_L}{k+1}rs^{(n+1)}c+\frac{k(1-w_L)}{k+1} rs^{(n+2)}c,
	\label{eq_piwalk}
\end{align}
which will later be useful for calculation.

To simplify, we assume a single initial cooperative player $1$ in our analysis, implying that $N_C=1$ and evolution favors cooperation if $\rho_C>1/N$. In this scenario, the condition for cooperation success under weak selection can be rewritten as per the equivalent form~\cite{nowak2010evolution} as shown in Eq.~(\ref{eq_successcondi}),
\begin{equation}\label{eq_successcondi}
	\left\langle\frac{\partial}{\partial\delta}(\mathcal{B}_1-\mathcal{D}_1)\right\rangle_{\begin{smallmatrix}\delta=0\\s_1=1\end{smallmatrix}}>0,
\end{equation}
where $\langle\cdot\rangle_{\begin{smallmatrix}\delta=0\\s_1=1\end{smallmatrix}}$ represents the expected value under neutral drift ($\delta=0$) and single cooperator ($s_1=1$). $\mathcal{B}_1$ is the probability of agent $1$ passing on its strategy to a neighbor. This occurs when a neighbor $i\in \Omega_1\setminus\{1\}$ of agent $1$ is randomly selected with a $1/N$ probability to update the strategy and learns agent $1$'s strategy with a $W(s_i\gets s_1)$ probability. In the same vein, $\mathcal{D}_1$ is the probability of agent $1$'s strategy being supplanted by a neighbor. This transpires when agent $1$ is randomly selected with a $1/N$ probability to update its strategy and learns the strategy of a neighbor $j\in \Omega_1\setminus\{1\}$ with a $W(s_1\gets s_j)$ probability. By applying Eq.~(\ref{eq_DB}) and $F_i=\exp{(\delta \pi_i)}$, we arrive at the equations summarized as follows:
\begin{subequations}\label{eq_bd}
	\begin{align}
		\mathcal{B}_1&=\sum_{i\in \Omega_1\setminus\{1\}}\frac{1}{N}W(s_i\gets s_1)
		=\sum_{i\in \Omega_1\setminus\{1\}}\frac{1}{N}
		\frac{(1-w_R)/k\cdot \exp{(\delta \pi_1)}}{w_R\exp{(\delta \pi_i)}+(1-w_R)/k\cdot \sum_{\ell\in \Omega_i\setminus\{i\}}\exp{(\delta \pi_{\ell})}},
	    \label{eq_b}
		\\
		\mathcal{D}_1&=\frac{1}{N}\sum_{j\in \Omega_1\setminus\{1\}}W(s_1\gets s_j)
		=\frac{1}{N}\sum_{j\in \Omega_1\setminus\{1\}}
        \frac{(1-w_R)/k\cdot \exp{(\delta \pi_j)}}{w_R\exp{(\delta \pi_1)}+(1-w_R)/k\cdot \sum_{\ell\in \Omega_1\setminus\{1\}}\exp{(\delta \pi_{\ell})}}.
	    \label{eq_d}
	\end{align}
\end{subequations}

In the further steps, we substitute Eq.~(\ref{eq_b}) and Eq.~(\ref{eq_d}) into Eq.~(\ref{eq_successcondi}) and compute it, as shown in Eq.~(\ref{eq_conditiondeduce}). 
\begin{align}\label{eq_conditiondeduce}
	&\left\langle\frac{\partial}{\partial\delta}(\mathcal{B}_1-\mathcal{D}_1)\right\rangle_{\begin{smallmatrix}\delta=0\\s_1=1\end{smallmatrix}}>0\nonumber
	\\
	\Leftrightarrow&~\frac{1-w_R}{Nk}
	\left(
	k\left\langle \pi_1\right\rangle_{\begin{smallmatrix}\delta=0\\s_1=1\end{smallmatrix}}
	-w_R\left\langle \sum_{i\in \Omega_1\setminus\{1\}}\pi_i\right\rangle_{\begin{smallmatrix}\delta=0\\s_1=1\end{smallmatrix}}
	-\frac{1-w_R}{k}\left\langle \sum_{i\in \Omega_1\setminus\{1\}}\sum_{\ell\in \Omega_i\setminus\{i\}}\pi_\ell\right\rangle_{\begin{smallmatrix}\delta=0\\s_1=1\end{smallmatrix}}
	\right)\nonumber
	\\
	&~-\frac{1-w_R}{Nk}
	\left(
	-kw_R \left\langle \pi_1\right\rangle_{\begin{smallmatrix}\delta=0\\s_1=1\end{smallmatrix}}
	+\left\langle \sum_{j\in \Omega_1\setminus\{1\}}\pi_j\right\rangle_{\begin{smallmatrix}\delta=0\\s_1=1\end{smallmatrix}}
        -(1-w_R)\left\langle \sum_{\ell\in \Omega_1\setminus\{1\}}\pi_{\ell}\right\rangle_{\begin{smallmatrix}\delta=0\\s_1=1\end{smallmatrix}}
	\right)>0\nonumber
	\\
	\Leftrightarrow&\left\langle\pi_1\right\rangle_{\begin{smallmatrix}\delta=0\\s_1=1\end{smallmatrix}}
	-\frac{2w_R}{k(1+w_R)}\left\langle \sum_{j\in \Omega_1\setminus\{1\}}\pi_j\right\rangle_{\begin{smallmatrix}\delta=0\\s_1=1\end{smallmatrix}}
	-\frac{1-w_R}{k^2(1+w_R)}\left\langle \sum_{i\in \Omega_1\setminus\{1\}}\sum_{\ell\in \Omega_i\setminus\{i\}}\pi_\ell\right\rangle_{\begin{smallmatrix}\delta=0\\s_1=1\end{smallmatrix}}>0\nonumber
	\\
	\Leftrightarrow&~
        \pi^{(0)}
	-\frac{2w_R}{1+w_R}\pi^{(1)}
	-\frac{1-w_R}{1+w_R}\pi^{(2)}>0.
\end{align}
Following the definition of random walks starting from agent $1$, we used Eq.~(\ref{eq_transtowalk}) in the last step of Eq.~(\ref{eq_conditiondeduce}).
\begin{equation}\label{eq_transtowalk}
    \pi^{(0)}=\left\langle\pi_1\right\rangle_{\begin{smallmatrix}\delta=0\\s_1=1\end{smallmatrix}},~
    \pi^{(1)}=\frac{1}{k}\left\langle \sum_{j\in \Omega_1\setminus\{1\}}\pi_j\right\rangle_{\begin{smallmatrix}\delta=0\\s_1=1\end{smallmatrix}},~
    \pi^{(2)}=\frac{1}{k^2}\left\langle \sum_{i\in \Omega_1\setminus\{1\}}\sum_{\ell\in \Omega_i\setminus\{i\}}\pi_\ell\right\rangle_{\begin{smallmatrix}\delta=0\\s_1=1\end{smallmatrix}}.
\end{equation}

To transform the strategy quantity $s^{(n)}$ into walk quantity $p^{(n)}$, the probability that one returns to the starting vertex after $n$-step random walks, we use the substitution in Eq.~(\ref{eq_stop}), as suggested by Allen and Nowak~\cite{allen2014games}:
\begin{equation}\label{eq_stop}
	s^{(n)}-s^{(n+1)}=\frac{\mu}{2}(Np^{(n)}-1)+\mathcal{O}(\mu^2),
\end{equation}
where $\mu\to 0^+$ is an auxiliary parameter, which will be eliminated later, and $\mathcal{O}(\mu^2)=0$. Based on Eq.~(\ref{eq_stop}), we can then further develop Eq.~(\ref{eq_stop2}):
\begin{align}\label{eq_stop2}
    s^{(n)}
    -\frac{2w_R}{1+w_R}s^{(n+1)}
    -\frac{1-w_R}{1+w_R}s^{(n+2)}
    &=(s^{(n)}-s^{(n+1)})
    +\frac{1-w_R}{1+w_R}(s^{(n+1)}-s^{(n+2)}) \nonumber\\
    &=\frac{\mu}{2}\left(Np^{(n)}+\frac{1-w_R}{1+w_R}Np^{(n+1)}-\frac{2}{1+w_R}\right)
    +\mathcal{O}(\mu^2).
\end{align}

Utilizing this, we can further calculate the condition for cooperation success as given by Eq.~(\ref{eq_conditiondeduce}). First, we use Eq.~(\ref{eq_piwalk}) to replace the payoff quantity $\pi^{(n)}$ with strategy quantity $s^{(n)}$. Second, we use Eq.~(\ref{eq_stop2}) to replace the strategy quantity $s^{(n)}$ with walk quantity $p^{(n)}$. This logic leads us to Eq.~(\ref{eq_calcu}):
\begin{align}\label{eq_calcu}
	&~\pi^{(0)}
	-\frac{2w_R}{1+w_R}\pi^{(1)}
	-\frac{1-w_R}{1+w_R}\pi^{(2)}>0\nonumber
	\\
	\Leftrightarrow&\left(\frac{w_L}{k+1}r-1\right)s^{(0)}c+\frac{1+(k-1)w_L}{k+1}rs^{(1)}c+\frac{k(1-w_L)}{k+1} rs^{(2)}c\nonumber\\
	&-\frac{2w_R}{1+w_R}
	\left\{\left(\frac{w_L}{k+1}r-1\right)s^{(1)}c+\frac{1+(k-1)w_L}{k+1}rs^{(2)}c+\frac{k(1-w_L)}{k+1} rs^{(3)}c\right\}\nonumber\\
	&-\frac{1-w_R}{1+w_R}
	\left\{\left(\frac{w_L}{k+1}r-1\right)s^{(2)}c+\frac{1+(k-1)w_L}{k+1}rs^{(3)}c+\frac{k(1-w_L)}{k+1} rs^{(4)}c\right\}>0\nonumber
	\\
	\Leftrightarrow&
	\left(\frac{w_L}{k+1}r-1\right)
    \left(s^{(0)}-\frac{2w_R}{1+w_R}s^{(1)}-\frac{1-w_R}{1+w_R}s^{(2)}\right)\nonumber\\
	&+\frac{1+(k-1)w_L}{k+1}r
    \left(s^{(1)}-\frac{2w_R}{1+w_R}s^{(2)}-\frac{1-w_R}{1+w_R}s^{(3)}\right)\nonumber\\
	&+\frac{k(1-w_L)}{k+1} r
    \left(s^{(2)}-\frac{2w_R}{1+w_R}s^{(3)}-\frac{1-w_R}{1+w_R}s^{(4)}\right)>0\nonumber
	\\
	\Leftrightarrow&
	\left(\frac{w_L}{k+1}r-1\right)
    \left(Np^{(0)}+\frac{1-w_R}{1+w_R}Np^{(1)}-\frac{2}{1+w_R}\right)\nonumber\\
	&+\frac{1+(k-1)w_L}{k+1}r
    \left(Np^{(1)}+\frac{1-w_R}{1+w_R}Np^{(2)}-\frac{2}{1+w_R}\right)\nonumber\\
	&+\frac{k(1-w_L)}{k+1} r
    \left(Np^{(2)}+\frac{1-w_R}{1+w_R}Np^{(3)}-\frac{2}{1+w_R}\right)>0.
\end{align}

The walk quantity $p^{(n)}$ can be directly perceived by analyzing the topology of the network structure. One remains in the starting vertex if not walking, so $p^{(0)}=1$. A single step cannot encompass leaving and returning to the starting vertex, hence $p^{(1)}=0$. On a square lattice, the probability that one returns to the starting vertex after two steps is $p^{(2)}=1/k$. Finally, the value of $p^{(3)}$ varies from case to case. In short, $p^{(3)}=0$ for von~Neumann neighborhood and $p^{(3)}=3/64$ for Moore neighborhood (for more details, refer to Ref.~\cite{wang2023inertia}).

By applying the previously mentioned values of $p^{(0)}=1$, $p^{(1)}=0$, and $p^{(2)}=1/k$, but retaining $p^{(3)}$, we can further calculate Eq.~(\ref{eq_calcu}) to reach the final result as shown in Eq.~(\ref{eq_calcu2}):
\begin{align}\label{eq_calcu2}
	&~\pi^{(0)}
	-\frac{2w_R}{1+w_R}\pi^{(1)}
	-\frac{1-w_R}{1+w_R}\pi^{(2)}>0\nonumber
	\\
	\Leftrightarrow&
	\left(\frac{w_L}{k+1}r-1\right)
    \left(N-\frac{2}{1+w_R}\right)
        +\frac{1+(k-1)w_L}{k+1}r\left(\frac{1-w_R}{1+w_R}\frac{N}{k}-\frac{2}{1+w_R}\right)\nonumber\\
	&+\frac{k(1-w_L)}{k+1} r\left(\frac{N}{k}+\frac{1-w_R}{1+w_R}Np^{(3)}-\frac{2}{1+w_R}\right)>0\nonumber
	\\
	\Leftrightarrow&~
	r>\frac{(N-2+N w_R)(G-1)G}
    {N(G-1)^2 (1-w_L)(1-w_R) p^{(3)}+N(G-2)(w_L-w_L w_R+w_R)+(N+2-2G)G}
	\equiv r^\star.
\end{align}
This provides the condition $r>r^\star$ for cooperation success. Notably, the critical synergy factor $r^\star$ is only a function of the population $N$, group size $G$, higher-order network structure $p^{(3)}$, self-allocation $w_L$, and updating inertia $w_R$.

Table~\ref{tb_value} summarizes the primary outcomes related to the critical synergy factor, $r^\star$, along with their corresponding large population limits ($N\to +\infty$), derived from taking specific parameters in Eq.~(\ref{eq_calcu2}). Following the convention in much of the prior literature, we consider the death--birth rule ($w_R=0$) as the benchmark scenario. In this context, we present the reduced $r^\star$ values corresponding to three distinct scenarios: equal allocation ($w_L=1/G$), allocation to other players ($w_L=0$), and allocation to the organizer ($w_L=1$). In addition, we explore a situation where the self-allocation and updating inertia are congruent ($w_L=w_R\equiv w$), leading to consistency in the self-loops of allocation and updating. The trajectories of this case in the $w_R$-$w_L$ parameter plane are visually represented in Fig.~\ref{fig_Diagram} for an intuitive understanding.

\begin{table*}[width=\linewidth,pos=h]
\caption{Critical synergy factors $r^\star$ for cooperation success under typical parameter values. All results are obtained by substituting specific parameter values into Eq.~(\ref{eq_calcu2}).}\label{tb_value}
\begin{tabular*}{\tblwidth}{@{} LL@{} }
\toprule
\multicolumn{1}{L}{Special parameter} & 
The critical $r>r^\star$ for cooperation success
\\\midrule
/ &
$\displaystyle{r^\star=
\frac{(N-2+N w_R)(G-1)G}
{N(G-1)^2 (1-w_L)(1-w_R) p^{(3)}+N(G-2)(w_L-w_L w_R+w_R)+(N+2-2G)G}}$
\\
$w_R=0$ & 
$\displaystyle{r^\star=
\frac{(N-2)(G-1)G}
{N(G-1)^2(1-w_L) p^{(3)}+N(G-2)w_L+(N+2-2G)G}}$
\rule{0em}{2em}\\
$w_R=0$, $\displaystyle{w_L=\frac{1}{G}}$ &
$\displaystyle{r^\star=
\frac{(N-2)G^2}
{N(G-1)^2 p^{(3)}+N(G+2)-2G^2}}$
\rule{0em}{2em}\\
$w_R=0$, $w_L=0$ &
$\displaystyle{r^\star=
\frac{(N-2)(G-1)G}{N(G-1)^2 p^{(3)}+(N+2-2G)G}}$
\rule{0em}{2em}\\
$w_R=0$, $w_L=1$ &
$\displaystyle{r^\star=
\frac{(N-2)G}{2(N-G)}}$
\rule{0em}{2em}\\
$w_R=w_L\equiv w$ & 
$\displaystyle{r^\star=
\frac{(N-2+Nw)(G-1)G}
{N(G-1)^2 (1-w)^2 p^{(3)}+N(G-2)(2-w)w+(N+2-2G)G}}$
\rule{0em}{2em}\\

$N\to +\infty$ & 
$\displaystyle{r^\star=
\frac{(1+w_R)(G-1)G}
{(G-1)^2 (1-w_L)(1-w_R) p^{(3)}+(G-2)(w_L-w_L w_R+w_R)+G}}$
\rule{0em}{2em}\\
$N\to +\infty$, $w_R=0$ & 
$\displaystyle{r^\star=
\frac{(G-1)G}
{(G-1)^2(1-w_L) p^{(3)}+(G-2)w_L+G}}$
\rule{0em}{2em}\\
$N\to +\infty$, $w_R=0$, $\displaystyle{w_L=\frac{1}{G}}$ &
$\displaystyle{r^\star=
\frac{G^2}
{(G-1)^2 p^{(3)}+G+2}}$
\rule{0em}{2em}\\
$N\to +\infty$, $w_R=0$, $w_L=0$ &
$\displaystyle{r^\star=
\frac{(G-1)G}{(G-1)^2 p^{(3)}+G}}$
\rule{0em}{2em}\\
$N\to +\infty$, $w_R=0$, $w_L=1$ &
$\displaystyle{r^\star=
\frac{G}{2}}$
\rule{0em}{2em}\\
$N\to +\infty$, $w_R=w_L\equiv w$ & 
$\displaystyle{r^\star=
\frac{(1+w)(G-1)G}
{(G-1)^2 (1-w)^2 p^{(3)}+(G-2)(2-w)w+G}}$
\rule{0em}{2em}\\
\bottomrule
\end{tabular*}
\end{table*}

Table~\ref{tb_value2} offers additional insights into the main outcomes associated with the critical synergy factor, $r^\star$, in relation to specific neighborhood types. We concentrate on two commonly used cases: von~Neumann neighborhood and Moore neighborhood. The former, von~Neumann neighborhood, lacks triangle motifs, resulting in $p^{(3)}=0$. Conversely, the latter, Moore neighborhood, is a rudimentary structure on a two-dimensional lattice that incorporates overlapping neighbors, yielding $p^{(3)}=3/64$~\cite{wang2023inertia}.

\begin{table}[width=\linewidth,pos=h]
\caption{Specified results of the critical synergy factor $r>r^\star$ for cooperation success on the square lattice with different group sizes. The cases of $G=5$ (von~Neumann neighborhood), $G=9$ (Moore neighborhood) are presented. See Ref.~\cite{wang2023inertia} for the visualization of $p^{(3)}$ value in each case.}\label{tb_value2}
\begin{tabular*}{\tblwidth}{@{} LLL@{} }
\toprule
\multicolumn{1}{L}{Special parameter} & 
$G=5$, $p^{(3)}=0$ & 
$G=9$, $p^{(3)}=3/64$
\\\midrule
/ &
$\displaystyle{r^\star=
\frac{20N w_R+20N-40}
{3N(w_L-w_L w_R+w_R)+5N-40}}$ &
$\displaystyle{r^\star=
\frac{18Nw_R+18N-36}
{N(w_L-w_L w_R+w_R)+3N-36}}$
\\
$w_R=0$ & 
$\displaystyle{r^\star=
\frac{20N-40}
{3Nw_L+5N-40}}$ &
$\displaystyle{r^\star=
\frac{18N-36}
{Nw_L+3N-36}}$
\rule{0em}{2em}\\
$w_R=0$, $\displaystyle{w_L=\frac{1}{G}}$ &
$\displaystyle{r^\star=
\frac{25N-50}
{7N-50}}$ &
$\displaystyle{r^\star=
\frac{81N-162}
{14N-162}}$
\rule{0em}{2em}\\
$w_R=0$, $w_L=0$ &
$\displaystyle{r^\star=
\frac{4N-8}{N-8}}$ &
$\displaystyle{r^\star=
\frac{6N-12}{N-12}}$
\rule{0em}{2em}\\
$w_R=0$, $w_L=1$ &
$\displaystyle{r^\star=
\frac{5N-10}{2N-10}}$ &
$\displaystyle{r^\star=
\frac{9N-18}{2N-18}}$
\rule{0em}{2em}\\
$w_R=w_L\equiv w$ & 
$\displaystyle{r^\star=
\frac{20Nw+20N-40}
{3N(2-w)w+5N-40}}$ &
$\displaystyle{r^\star=
\frac{18Nw+18N-36}
{N(2-w)w+3N-36}}$
\rule{0em}{2em}\\

$N\to +\infty$ & 
$\displaystyle{r^\star=
\frac{20w_R+20}
{3(w_L-w_L w_R+w_R)+5}}$ &
$\displaystyle{r^\star=
\frac{18w_R+18}
{w_L-w_L w_R+w_R+3}}$
\rule{0em}{2em}\\
$N\to +\infty$, $w_R=0$ & 
$\displaystyle{r^\star=
\frac{20}
{3w_L+5}}$ &
$\displaystyle{r^\star=
\frac{18}
{w_L+3}}$
\rule{0em}{2em}\\
$N\to +\infty$, $w_R=0$, $\displaystyle{w_L=\frac{1}{G}}$ &
$\displaystyle{r^\star=
\frac{25}
{7}\approx 3.5714}$ &
$\displaystyle{r^\star=
\frac{81}
{14}}\approx 5.7857$
\rule{0em}{2em}\\
$N\to +\infty$, $w_R=0$, $w_L=0$ &
$\displaystyle{r^\star=
4}$ &
$\displaystyle{r^\star=
6}$
\rule{0em}{2em}\\
$N\to +\infty$, $w_R=0$, $w_L=1$ &
$\displaystyle{r^\star=
\frac{5}{2}=2.5}$ &
$\displaystyle{r^\star=
\frac{9}
{2}=4.5}$
\rule{0em}{2em}\\
$N\to +\infty$, $w_R=w_L\equiv w$ & 
$\displaystyle{r^\star=
\frac{20w+20}
{3(2-w)w+5}}$ &
$\displaystyle{r^\star=
\frac{18w+18}
{(2-w)w+3}}$
\rule{0em}{2em}\\
\bottomrule
\end{tabular*}
\end{table}

\subsection{The conflict between self-allocation and self-confidence}\label{sec_conflict}
Utilizing the analytical expression of the critical synergy factor $r^\star$, we can examine the combined impact of self-allocation $w_L$ and self-confidence $w_R$ on cooperation. From an intuitive perspective, a decrease in the $r^\star$ value needed for cooperation success (i.e., $r>r^\star$) fosters cooperation.

\begin{figure}
	\centering
		\includegraphics[width=.75\textwidth]{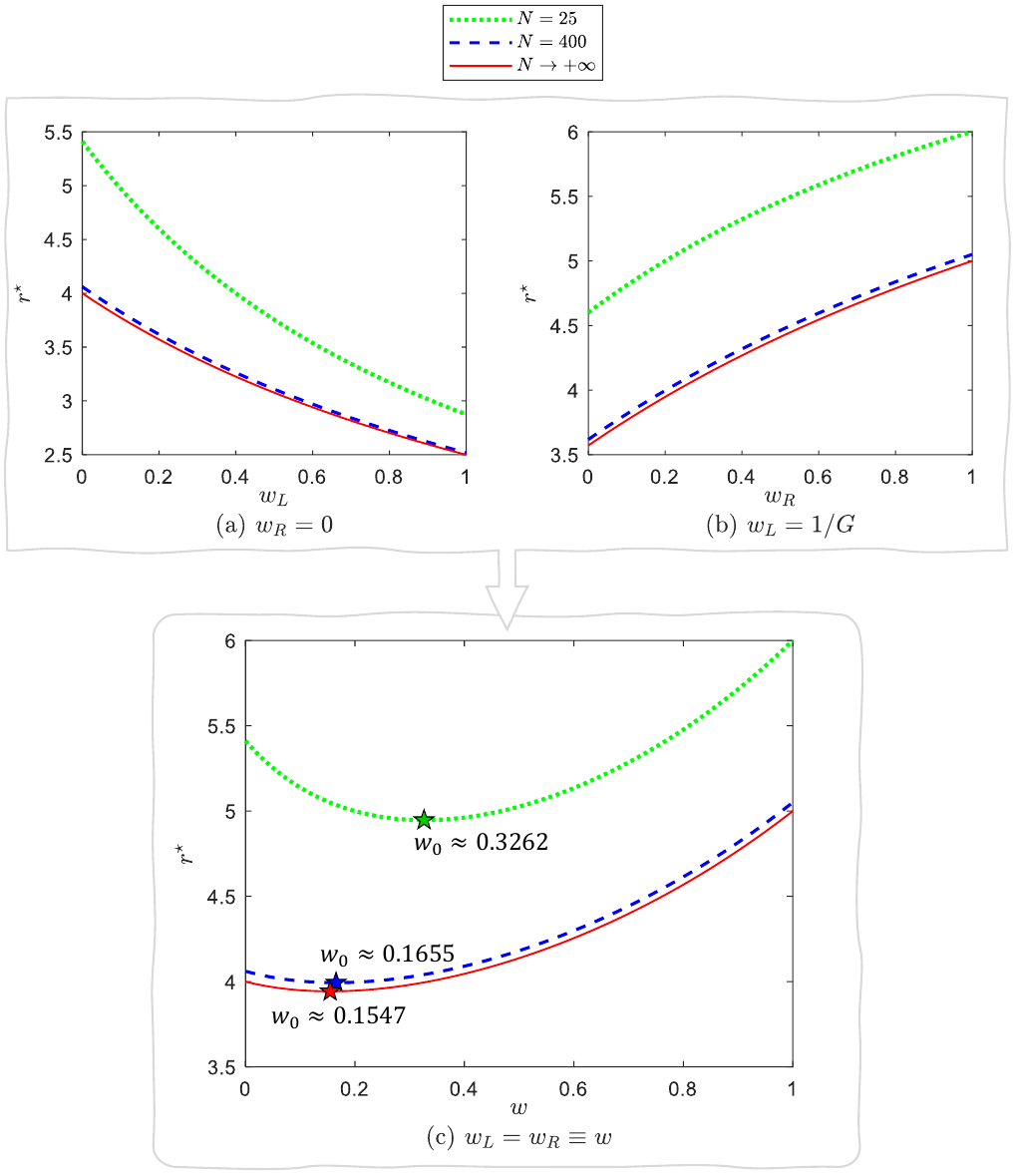}
	\caption{(a) The critical synergy factor $r^\star$ as a function of the self-allocation weight $w_L$ with $w_R=0$. The implementation of biased payoff allocation consistently fosters cooperation when the updating inertia $w_R$ is held constant. (b) The critical synergy factor $r^\star$ as a function of strategy updating inertia $w_R$ with $w_L=1/G$. When considered in isolation, the updating inertia inhibits cooperation. (c) The critical synergy factor $r^\star$ as a function of $w\equiv w_L=w_R$. As both effects operate in conjunction, the concurrent increase in weight factors initially promotes, then later hinders cooperation. The star indicates the optimal $w=w_0$ value where the cooperation-enhancing effect reaches its peak. Other parameters include $G=5$ and $p^{(3)}=0$.} 
	\label{fig_1D}
\end{figure}

By referring to Eq.~(\ref{eq_calcu2}), we can confirm that $\partial r^\star/\partial w_L<0$ holds for the specified neighborhood types. This indicates that an increase in self-allocation diminishes $r^\star$ and thereby enhances cooperation. Fig.~\ref{fig_1D}(a) portrays the critical synergy factor $r^\star$ as a function of self-allocation $w_L$ for von Neumann neighborhood under the condition of death--birth updating ($w_R=0$). Regardless of the population size, directing the public goods towards the organizer invariably stimulates cooperation. 

Similarly, we find $\partial r^\star/\partial w_R>0$ for the designated neighborhood types. This suggests that an increase in self-confidence, or alternatively, an increase in updating inertia, acts to obstruct cooperation. This effect aligns with observations made in simpler models by prior studies~\cite{wang2023evolution,wang2023inertia,wang2023conflict}. With the von~Neumann neighborhood and $w_L=1/G$, the critical synergy factor $r^\star$ as a function of updating inertia is depicted in Fig.~\ref{fig_1D}(b). Across varying population sizes, an increase in updating inertia consistently hampers cooperation.

The aforementioned observations create a fascinating dynamic when both effects coexist. Specifically, the divergent outcomes of biased allocation and self-confidence pose a question: how does the system respond when we enhance the weights of these factors simultaneously? Does it stimulate or inhibit cooperation? To explore this, we set $w_L=w_R\equiv w$ and illustrate the critical synergy factor $r^\star$ as a function of $w$ in Fig.~\ref{fig_1D}(c). The figure reveals that an initial increase in the self-loop of allocation and strategy updating fosters cooperation, but once the weight surpasses a certain level, this effect reverses, ultimately discouraging cooperation. There exists an optimal self-loop weight $w_0$, which minimizes the $r^\star$ value and is thus most beneficial for cooperation. We can derive the analytical expression for this optimal self-loop value by solving $\partial r^\star/\partial w=0$. The solution is given as:
\begin{equation}\label{eq_w0}
    w_0=\frac{1}{N}\left(
    -(N-2)+\sqrt{2} \sqrt{2(N-1)^2+\frac{N(N-G)(G-1)}{(G-1)^2 p^{(3)}-G+2}}\right),
\end{equation}
which is a function of population size $N$, group size $G$, and the higher-order network structure $p^{(3)}$. This weight level provides the most favorable condition for the evolution of cooperation.

By setting $N\to +\infty$ in Eq.~(\ref{eq_w0}), we obtain the large population limit of $w_0$ as:
\begin{equation}
    w_0=-1+\sqrt{2} \sqrt{2+\frac{G-1}{(G-1)^2 p^{(3)}-G+2}}.
\end{equation}

\begin{figure}
	\centering
		\includegraphics[width=\textwidth]{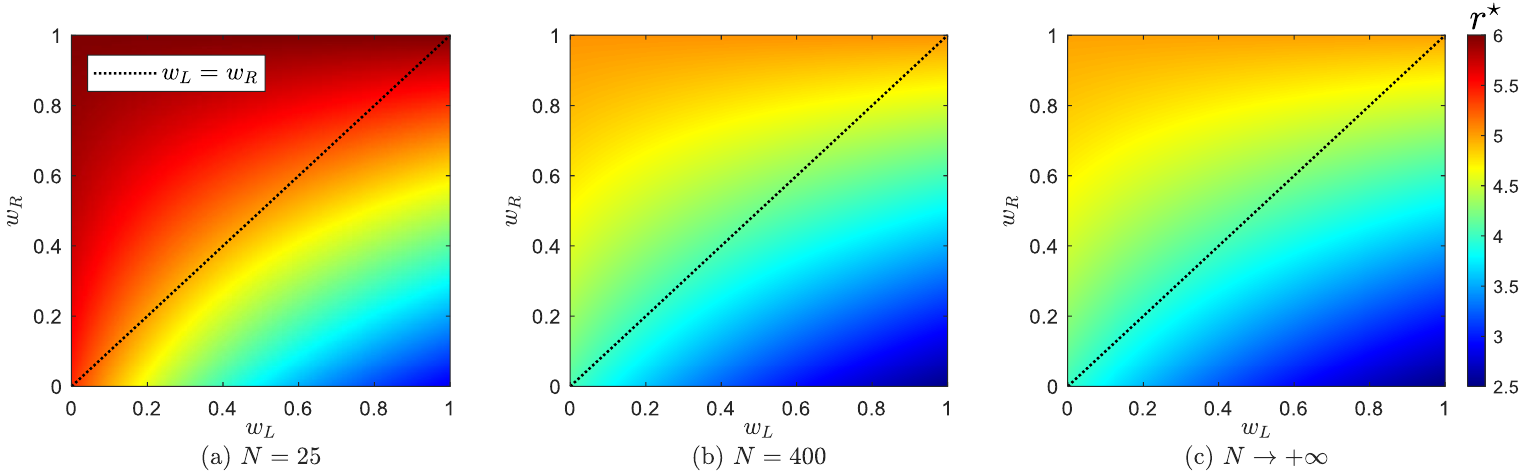}
	\caption{The color-coded critical synergy factor $r^\star$ across the comprehensive $w_R$-$w_L$ parameter plane. The different panels display results for various system sizes: (a) $N=25$, (b) $N=400$, and (c) $N\to+\infty$. The dotted diagonal line indicates the trajectory utilized in Fig.~\ref{fig_1D}(c). The other parameters are $G=5$ and $p^{(3)}=0$.} 
	\label{fig_2D}
\end{figure}

To provide a broader perspective on the simultaneous influences of these factors, we introduce a heat map of the critical synergy factor $r^\star$ across the complete $w_R$-$w_L$ parameter plane in Fig.~\ref{fig_2D}. The diagonal dotted line within the figure represents the trajectory discussed in Fig.~\ref{fig_1D}(c). This plot reveals certain general characteristics regarding the collective impact of self-loop effects. Specifically, the immediate effect of biased payoff allocation on the critical synergy factor is more pronounced when $w_R$ is small, whereas the $w_R$ dependency of $r^\star$ is moderate for large $w_R$ values. The inverse is true when considering the $w_R$ dependency of $r^\star$, as it changes more dramatically when $w_L$ is low, while the $w_R$ dependency remains moderate for small $w_L$ values.

When maintaining the aforementioned diagonal trajectory, we can identify some general trends regarding the $w$-dependence. Specifically, we can confirm that the $r^\star$ value at $w=0$ is consistently lower than the one at $w=1$, that is, $\left.r^\star\right|_{w=0}<\left.r^\star\right|_{w=1}$. Applying $w=1$ and $w=0$ in Eq.~(\ref{eq_calcu2}), we find $\left.r^\star\right|_{w=1}=(N-1)G/(N-G)$ and $\left.r^\star\right|_{w=0}=(N-2)(G-1)G/[N(G-1)^2 p^{(3)}+(N+2-2G)G]$, respectively. Given that $N(G-1)^2 p^{(3)}>0$ always stands, we deduce $\left.r^\star\right|_{w=0}<(N-2)(G-1)G/[(N+2-2G)G]=[(N-1)G-(G-2)-N]/[N-G-(G-2)]$. And since $(N-1)G>N-G$ and $-(G-2)<0$, it follows that $\left.r^\star\right|_{w=0}<[(N-1)G-N]/(N-G)<(N-1)G/(N-G)$. Therefore, $\left.r^\star\right|_{w=0}<\left.r^\star\right|_{w=1}$ always holds true. This indicates that, on a larger scale, when both self-loop effects are significant, the outcome is dominated by the impact of self-confidence, which hinders cooperation. This effect is more pronounced in a topology containing triangle motifs, such as the Moore neighborhood where each player forms a $G=9$-member group with overlapping neighbors. This case is discussed in more detail in Appendix~\ref{sec_appen}.

\section{Numerical simulation}\label{sec_nume}
To validate our theoretical analysis, we performed Monte Carlo simulations. Initially, each agent is randomly assigned either cooperation or defection, such that $N_C\approx N/2$. Consequently, as outlined at the beginning of Section~\ref{sec_theo}, evolution favors cooperation if $\rho_C>1/2$. To compute the expected cooperation level $\rho_C$, we permit up to $40,000$ full Monte Carlo steps per run (if all agents become either cooperators or defectors, that specific run may be terminated earlier), and record the cooperation proportion at the last step as the result of each run. The expected cooperation level $\rho_C$ is then the average across multiple independent runs. Based on our empirical exploration, for $N=25$, $\rho_C$ is the average over $1,000,000$ runs; for $N=400$, $\rho_C$ is the average over $10,000$ runs; for $N=10000$, $\rho_C$ is obtained from a single run.

\begin{figure}
\centering
    \includegraphics[width=\textwidth]{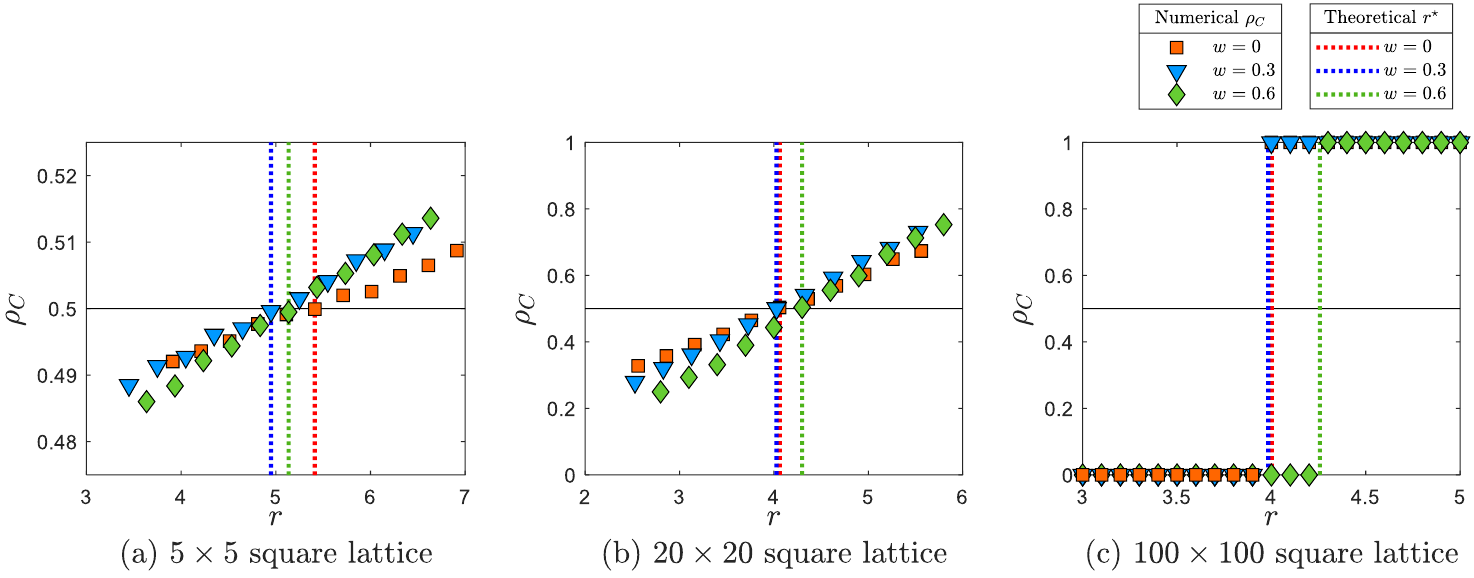}
\caption{Monte Carlo simulations validate the theoretical analysis in the public goods game incorporating self-allocation and updating inertia. The cooperative level $\rho_C$ is obtained from the simulations, as described in Sections~\ref{sec_model} and \ref{sec_nume}. The theoretical threshold for cooperative success, $r^\star$, is determined using Eq.~(\ref{eq_calcu2}). The parameters used are $w_L=w_R\equiv w$; $G=5$; $p^{(3)}=0$; $c=1$; with $\delta=0.01$ applied for panels~(a)-(b) and $\delta=0.1$ for panel~(c).} \label{fig_G5}
\end{figure}

Using the von~Neumann neighborhood, Fig.~\ref{fig_G5} illustrates the expected cooperation level $\rho_C$ as a function of the synergy factor $r$ at $w=0$, $w=0.3$, and $w=0.6$. In Fig.~\ref{fig_G5}(a), where $N=25$, substituting all parameter values into Eq.~(\ref{eq_calcu2}) gives $r^\star=5.4118$, $4.9493$, $5.1351$ for $w=0$, $0.3$, and $0.6$, respectively. Similarly, in Fig.~\ref{fig_G5}(b), for $N=400$, we get $r^\star=4.0612$, $4.0280$, $4.2992$. In Fig.~\ref{fig_G5}(c), where $N=10000$, we obtain $r^\star=4.0024$, $3.9835$, $4.2571$. As can be observed, the cooperation level $\rho_C$ rises with an increase in the synergy factor $r$, and $\rho_C>0.5$ when $r>r^\star$, thus affirming the theoretical analysis.

\section{Conclusion}\label{sec_conclu}
Collaborating on a project does not necessarily equate to equal benefits from the resulting income. For instance, an individual acting as the organizer of a group may allocate a different proportion of public goods to themselves than to other participants. If everyone follows the same protocol, allocating more public goods to the organizer boosts the gains in the game managed by oneself, but simultaneously leads to fewer gains in games organized by neighbors. Consequently, the impact of biased allocation on the level of cooperation is far from a simple question. Prior studies have demonstrated that this seemingly strategy-neutral mechanism actually promotes cooperation by preventing the diffusion of public goods~\cite{allen2013spatial,su2018understanding}.

On the other hand, if an individual allocates more public goods to themselves as an organizer, this attitude might also imply that the individual is more authoritative and confident, and less inclined to change their current strategy. Past observations have revealed that this inertia in strategy updating inhibits cooperation by slowing the aggregation of cooperators~\cite{wang2023evolution,wang2023inertia}. Thus, it can be concluded that biased allocation and strategy updating inertia play opposing roles in the evolution of cooperation.

Assuming that the measure of biased allocation and updating inertia are interconnected, this study focuses on their simultaneous presence and explores how they jointly influence cooperation. We derive a theoretical solution on a two-dimensional square lattice and identify the critical synergy factor $r^\star$ required for cooperation success. Consequently, cooperators are more likely to dominate when $r>r^\star$. Our primary interest lies in how $r^\star$ fluctuates on the plane of weight factors, which determine biased allocation and the extent of strategy updating inertia. Upon introducing the self-loop $w$ of allocation and updating, it initially promotes and later, for larger $w$ values, inhibits cooperation. In this scenario, we can identify an optimal self-loop value $w_0$ that is most conducive to cooperation. In other cases, where the network topology contains triangle motifs, the impact of strategy inertia is more potent, thus increasing the self-loop $w$ tends to hamper cooperation.

Moreover, we theoretically demonstrate that the cooperation threshold at $w=0$ is always smaller than at $w=1$. This suggests that the inhibitory effect of self-confidence on cooperation generally outweighs the facilitative effect of self-allocation on cooperation when the allocation and updating self-loop $w$ takes extreme values. These observations propose that although biased allocation may appear as an unfair protocol, its impact on cooperation is decidedly not detrimental. However, the self-confidence driven strategy updating inertia is always harmful, and cannot be offset by the effect of allocation.

\printcredits

\section*{Acknowledgement}
A.S. was supported by the National Research, Development and Innovation Office (NKFIH) under Grant No. K142948.

\appendix
\renewcommand\thefigure{\Alph{section}\arabic{figure}} 
\section{Moore neighborhood}\label{sec_appen}
\setcounter{figure}{0}
Our primary results are summarized in Eq.~(\ref{eq_calcu2}). It proposes that topology slightly influences the critical synergy factor $r^\star$ through the parameter $G$. However, a more complex consequence is embodied in the value of $p^{(3)}$. This factor creates a stark distinction between the von~Neumann and Moore neighborhoods, regardless of using the same vertex-transitive square lattice. For the von~Neumann neighborhood, the three-step quantity $p^{(3)}=0$, as there is no triangle motif. To explore the consequences of a non-zero $p^{(3)}$, we examine the Moore neighborhood, the simplest two-dimensional lattice that contains higher-order structure where $p^{(3)}=3/64$~\cite{wang2023inertia}.

\begin{figure}
	\centering
		\includegraphics[width=.75\textwidth]{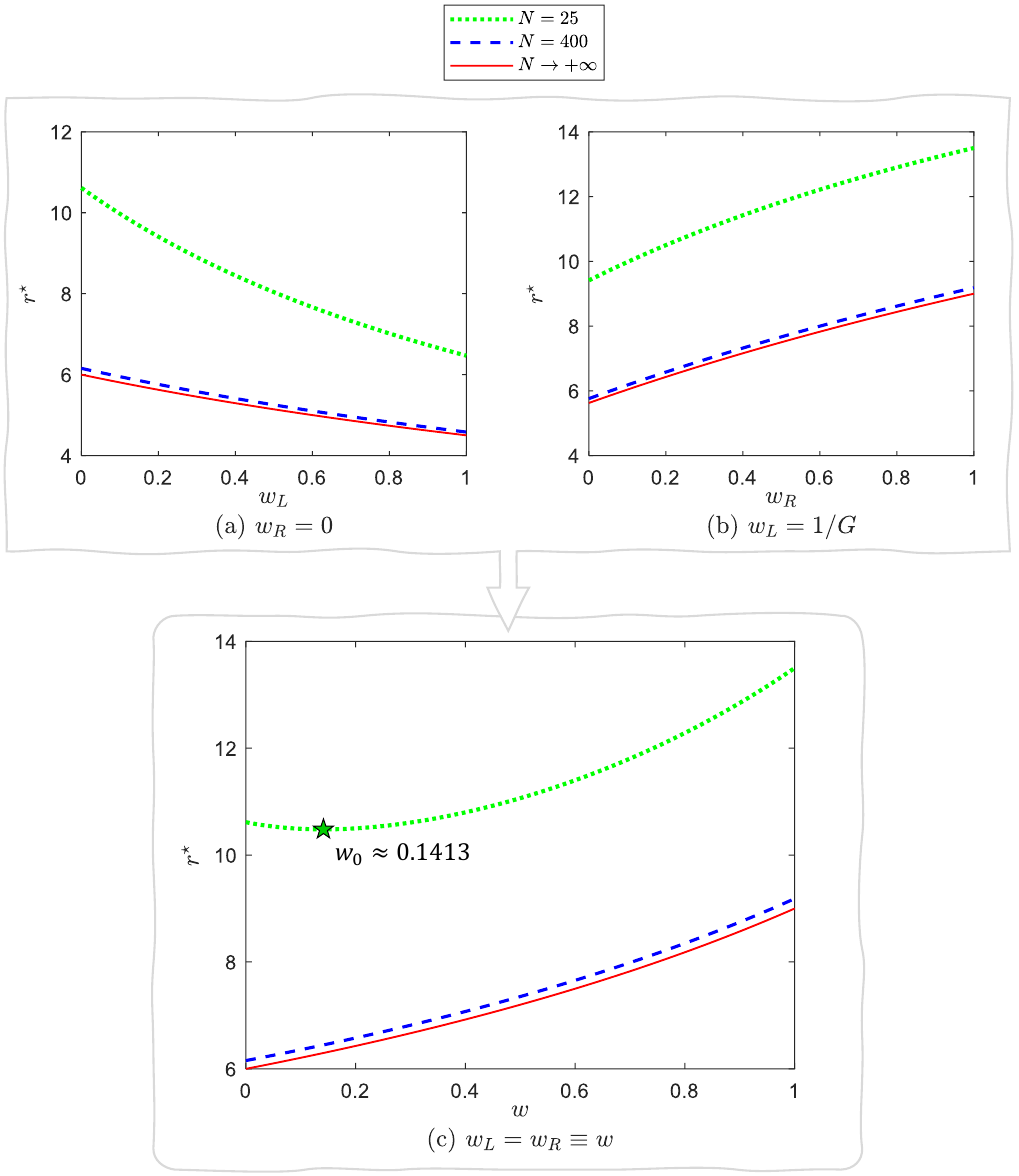}
	\caption{(a) The critical synergy factor $r^\star$ as a function of self-allocation weight $w_L$ when $w_R=0$. The introduction of biased payoff allocation consistently enhances cooperation when the other key parameter is held constant. (b) The critical synergy factor $r^\star$ as a function of strategy updating inertia $w_R$ when $w_L=1/G$. This single effect inhibits cooperation. (c) The critical synergy factor $r^\star$ as a function of $w\equiv w_L=w_R$. When both effects are operational, the concurrent increase of weight factors initially supports, but later hinders, cooperation for smaller system sizes. The star symbol indicates the optimal $w=w_0$ level where the cooperation-supporting effect is at its strongest. When the system size is sufficiently large, this effect disappears because the cooperation-weakening outcome of self-leaning is consistently stronger at this interaction topology. The other parameters are $G=9$, $p^{(3)}=3/64$.} 
	\label{fig_1D_Moore}
\end{figure}

The first two panels of Fig.~\ref{fig_1D_Moore} confirm that the separate impacts of biased allocation and strategy updating inertia are similar to those observed for the von~Neumann neighborhood. However, their combined influence on $r^\star$ diverges from the previous observation, as the self-confidence-based inertia is significantly stronger in this context, making the increase of the mutual weight factor $w$ detrimental to the success of cooperation.

\begin{figure}
	\centering
		\includegraphics[width=\textwidth]{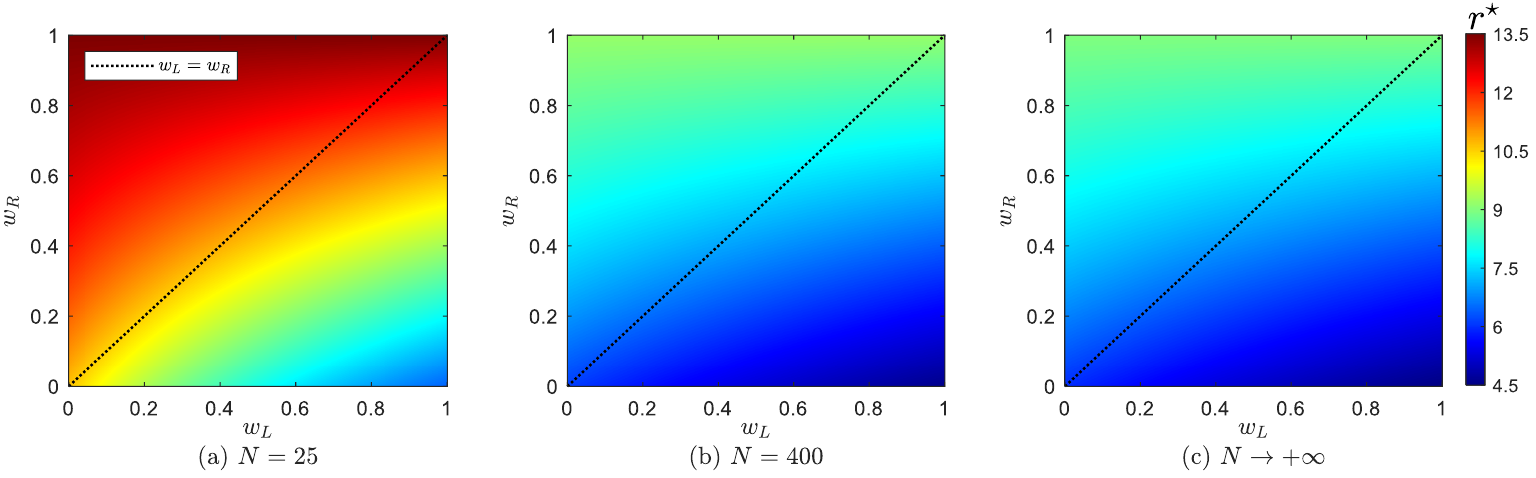}
	\caption{The color-coded values of the critical synergy factor $r^\star$ across the complete $w_R$-$w_L$ parameter plane. As indicated, the panels display results for different system sizes: (a) $N=25$, (b) $N=400$, and (c) $N\to+\infty$. The diagonal dotted line denotes the trajectory used for Fig.~\ref{fig_1D_Moore}(c). The other parameters are $G=9$, $p^{(3)}=3/64$.} 
	\label{fig_2D_Moore}
\end{figure}

This effect is generally valid and becomes evident when we compare the color-coded heat map of the critical synergy factor $r^\star$ on the $w_R$-$w_L$ parameter plane. The main difference between the last panels of Fig.~\ref{fig_2D} and Fig.~\ref{fig_2D_Moore} is the minimal change in the value of $r^\star$ as we move horizontally on the parameter plane of Fig.~\ref{fig_2D}(c). This suggests that changes in $w_L$ have only a minimal impact on cooperation, because the value of $w_R$ is the determining factor here.

\begin{figure}
\centering
    \includegraphics[width=\textwidth]{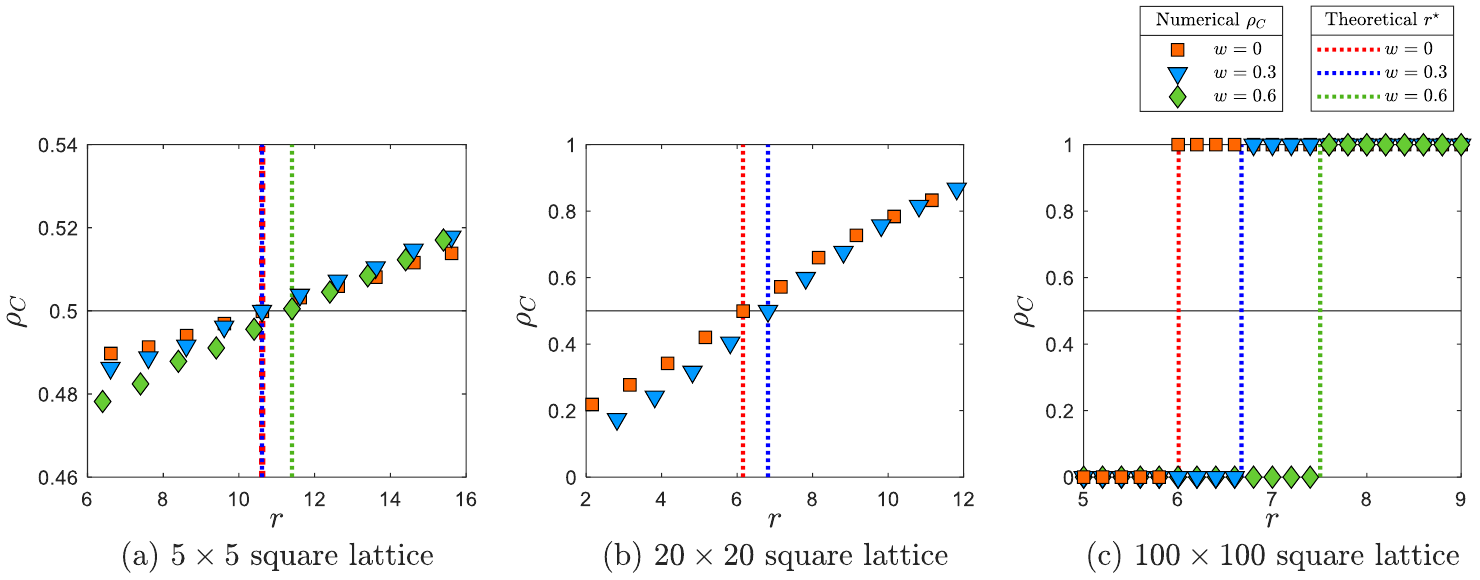}
\caption{The Monte Carlo simulations for the public goods game incorporating self-allocation and updating inertia support our theoretical analysis. The numerical cooperation level $\rho_C$ is derived from the Monte Carlo simulation as outlined in Sections~\ref{sec_model} and \ref{sec_nume}. The theoretical cooperation success threshold $r^\star$ is determined by Eq.~(\ref{eq_calcu2}). Parameters: $w_L=w_R\equiv w$; $G=5$; $p^{(3)}=0$; $c=1$; $\delta=0.01$ for panels~(a)-(b), and $\delta=0.1$ for panel~(c).} \label{fig_G9}
\end{figure}

Our final Fig.~\ref{fig_G9} presents a comparison of the results from our analytical and numerical calculations. In Fig.~\ref{fig_G9}(a), where $N=25$, substituting all parameter values into Eq.~(\ref{eq_calcu2}) yields $r^\star=10.6154$, $10.6087$, $11.4000$ for $w=0$, $0.3$, and $0.6$, respectively. Similarly, in Fig.~\ref{fig_G9}(b), for $N=400$, we obtain $r^\star=6.1546$, $6.8158$ for $w=0$, $0.3$. In Fig.~\ref{fig_G9}(c), where $N=10000$, we calculate $r^\star=6.0060$, $6.6725$, $7.5061$ for $w=0$, $0.3$, and $0.6$. As before, the simulations confirm our theoretical predictions well.

% %% Loading bibliography style file
% % \bibliographystyle{model1-num-names}
% \bibliographystyle{elsarticle-num-names}
% \bibliography{cas-refs}

\end{document}